\documentclass[twocolumn]{svjour3}          
\smartqed  
\usepackage{graphicx}
\begin{document}

\title{Flat band in topological matter 
}
\subtitle{possible route to room-temperature superconductivity}


\author{G.E. Volovik 
}


\institute{G.E. Volovik \at
           Low Temperature Laboratory, Aalto University, P.O. Box 15100, FI-00076 AALTO, Finland \\
and\\
L.D. Landau Institute for Theoretical Physics, Kosygina 2, 119334 Moscow, Russia \\
              Tel.: +358 9 470 22963\\
              Fax: +358 9 470 22969\\
              \email{volovik@ltl.tkk.fi}           
 }

\date{Received: date / Accepted: date}

\maketitle

\begin{abstract}
 Topological media are systems whose properties are protected by
  topology and thus are robust to deformations of the system.  In
  topological insulators and superconductors the bulk-surface and
  bulk-vortex correspondence gives rise to the gapless Weyl, Dirac or
  Majorana fermions on the surface of the system and inside vortex
  cores. In gapless topological media, the
  bulk-surface and bulk-vortex correspondence 
  produce topologically protected gapless fermions without dispersion
  -- the flat band. Fermion zero modes forming the flat band are
  localized on the surface of topological media
  with protected nodal lines and in the vortex core
  in systems with topologically protected Fermi points (Weyl points). Flat band has an extremely singular density of
  states, and this property may give rise in particular
  to surface superconductivity which in principle could exist even at room temperature.

\end{abstract}

\begin{figure*}
\centering
  \includegraphics[width=\textwidth]{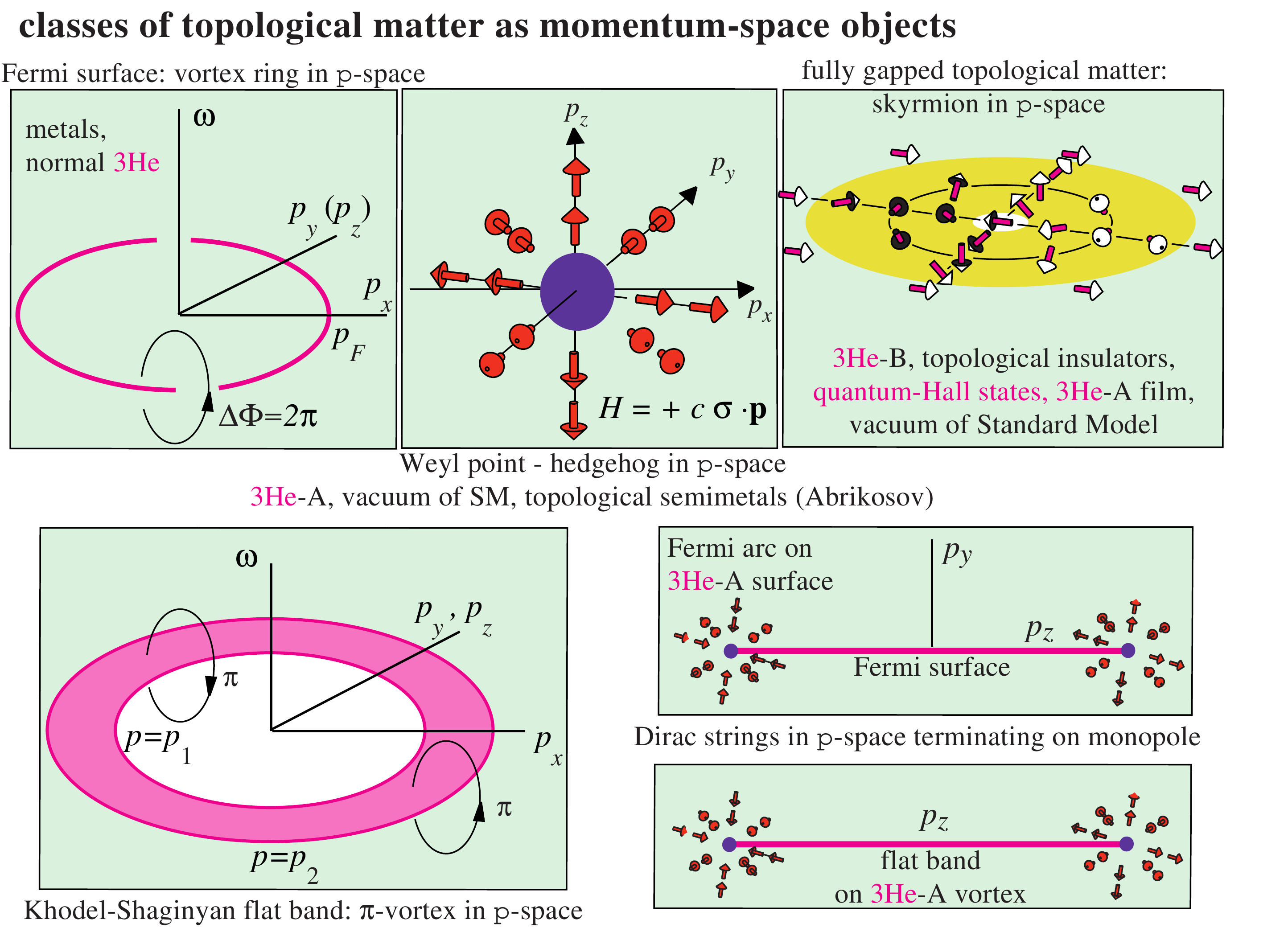}
\caption{
Topological matter, represented in terms of topological objects in momentum space.
({\it top left}): Fermi surface is the momentum-space analog of the vortex line: the phase of the Green's function changes by $2\pi N_1$ around the element of the line in $(\omega, {\mathbf p})$-space. ({\it top middle}):  Fermi point (Weyl point) is the counterpart of a hedgehog and a magnetic monopole. The hedgehog in this figure has integer topological charge $N_3=+1$, and close to this Fermi point the fermionic quasiparticles behave as Weyl fermions. Nontrivial topological charges $N_1$ and $N_3$ in terms of Green's functions support the stability of the Fermi surfaces and Weyl points with respect to perturbations including interactions \cite{Volovik2003,EssinGurarie2011}. ({\it top right}):  Topological insulators and fully gapped topological superfluids/superconductors are textures in momentum space: they have no singularities in the Green's function and thus no nodes in the energy spectrum in the bulk. This figure shows a skyrmion in the two-dimensional momentum space, which characterizes two-dimensional topological insulators exhibiting intrinsic quantum Hall or spin-Hall effect
\cite{Volovik2003}. ({\it bottom left}):  Flat band emerging in strongly interacting systems  \cite{Khodel1990}. This dispersionless Fermi band is analogous to a soliton terminated by half-quantum vortices: the phase of the Green's function changes by $\pi$ around the edge of the flat  band   \cite{NewClass}.  
Topologically protected flat band emerges on the surface of materials with
nodal lines in bulk, see Fig. \ref{SpiralFlatBand} and Refs. \cite{SchnyderRyu2010,Heikkila2011}. 
({\it bottom right}): Fermi arc on the surface of $^3$He-A \cite{Tsutsumi2011} and of topological semi-metals with Weyl points \cite{XiangangWan2011,Burkov2011}  serves as the momentum-space analog of a Dirac string terminating on a monopole. The Fermi surface formed by the surface bound states terminates on the points where the spectrum of zero energy states merge with the continuous spectrum in the bulk, i.e. with the Weyl points. The nodal topological systems with Weyl points also demonstrate the bulk-vortex correspondence. The core of the $^3$He-A vortex contains the topologically protected one-dimensional flat band which terminates on Weyl points \cite{Volovik2010}.
}
\label{classes}       
\end{figure*}

Discovery of topological insulators and graphene gave new impulse to investigation of topological media, which started after discovery of topological phases of superfluid $^3$He in seventies. Many quantum condensed matter systems are strongly correlated and strongly interacting fermionic systems, which cannot be treated perturbatively. However, topology allows us to determine generic features of their fermionic spectrum, which are robust to perturbation and interaction. Topological matter is characterized by a nontrivial topology in momentum space. The momentum-space topological invariants are in many respects similar to the real-space invariants, which describe topological defects in condensed matter systems, and cosmic strings and magnetic monopoles in particle physics 
(see Fig. \ref{classes}). In particular, the Fermi surface in metals is topologically stable, because it is analogous to the real-space vortex line in superfluids and superconductors. Its topological charge � the winding number -- cannot continuously change from 1 to 0. This makes the Fermi surface robust to perturbative interactions and is actually in the origin of the Landau theory of Fermi liquid, which is the effective low-energy theory of the systems with Fermi surface.

In the same way, the Fermi point in the energy spectrum (Dirac or Weyl point) is the analog of the real-space point defects, such as hedgehog in ferromagnets and magnetic monopole in particle physics. 
Different Fermi points may collide,
annihilate, and split again,
but their total topological charge is conserved in the same way as topologically-charged 't Hooft--Polyakov magnetic monopoles
in real space. The splitting or recombination of Fermi point in momentum space represents
an example of topological quantum phase transitions. 
Examples of systems with Weyl points are superfluid $^3$He in phase A, topological semimetals first discussed by Abrikosov and Beneslavskii in 1971 \cite{AbrikosovBeneslavskii1971}  and the vacuum of Standard Model of particle physics. The effective theories describing these systems at low energy are the theory of relativistic quantum fields and gravity, which emerge in the vicinity of the Weyl point. 
In Standard Model,
either Fermi points with opposite topological invariants annihilate each other, giving rise
to a Dirac mass (the process commonly known as Higgs mechanism),
or these Fermi points do not annihilate but split in momentum space,
giving rise to Lorentz violation presumably in neutrino sector~\cite{KlinkhamerVolovikIJMPA2005,KlinkhamerVolovik2011}.

\begin{figure}
  \includegraphics[width=1.0\textwidth]{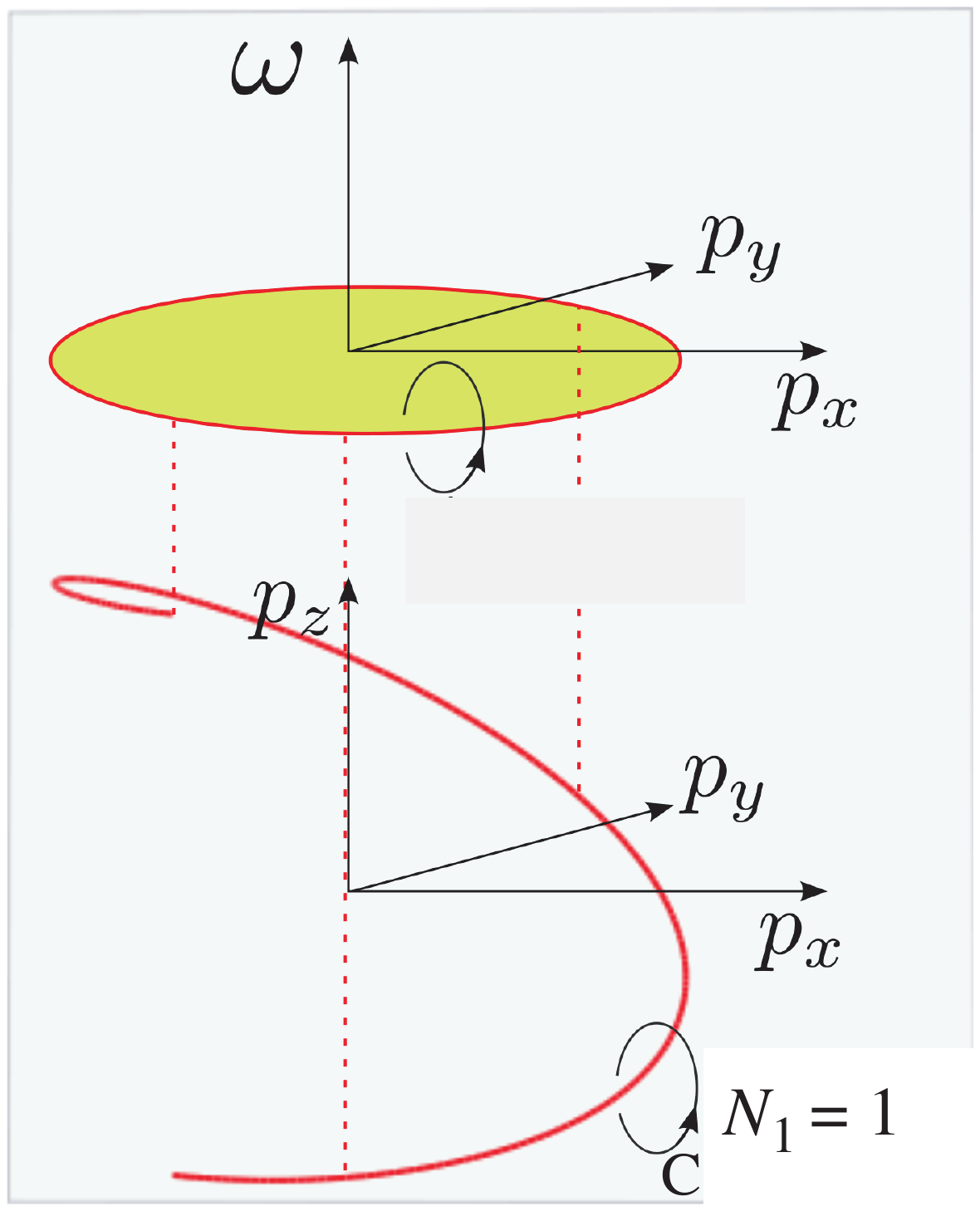}
\caption{The nodal line in the 3D topological semimetal gives rise to topologically protected bound states on the surface of the material. The energy spectrum of these bound states form the flat band which terminates on the projection of the nodal line to the surface, all the states within the flat band have zero energy.
Flat band emerging on the surface has an extremely singular density of states. In Ref. \cite{Heikkila2011} the special scenario of formation of the surface flat band has been considered, when it is obtained by stacking of $N$ graphene-like layers in the limit $N\rightarrow \infty$, if some discrete $Z_2$ symmetry is obeyed. For $N>2$ the singular density of states (DoS) takes place, $\nu_N(\epsilon)\sim \epsilon^{2/N -1}$. For $N=4$ layers one comes to $\nu(\epsilon)\sim \epsilon^{-1/2}$, which corresponds to the DoS at
the edge of a one-dimensional band. The latter for a long time has been
discussed as a possible source of enhancement of the
superconducting transition temperature, see e.g. \cite{Kopaev1987}.
Finally, in the limit of large number of layers the flat band with 
DoS $\nu(\epsilon)\sim \epsilon^{-1}$ is formed on the top and bottom surfaces of the material}
\label{SpiralFlatBand}       
\end{figure}

The fully gapped topological matter -- topological insulators and fully gapped topological superfluids such as superfluid $^3$He-B -- have no nodes in their bulk spectrum or any other singularities in momentum space. These systems are analogs of non-singular objects in real space -- coreless vortices, textures and skyrmions. The fully gapped systems, which have nonzero value of topological invariant in bulk, have gapless fermions on the boundary and in the core of quantized vortices. This relation between the bulk and edge properties is called the bulk-surface and bulk-vortex correspondence. In some systems the edge states or/and bound states in the vortex core have Majorana nature. 
The first discussion of the topological insulators can be found in Refs. \cite{Volkov1981,VolkovPankratov1985}; recent reviews are in Refs.  \cite{HasanKane2010,Xiao-LiangQi2011}. Exotic properties of the surface of topological insulators and fully gapped topological superfluids and superconductors can be found in 
Refs. \cite{Nomura2012,Stone2012}.

The nodal topological systems with Weyl fermions also demonstrate the bulk-surface and the bulk-vortex correspondence.  Due to bulk-vortex correspondence, the core of the $^3$He-A vortex contains the dispersionless branch of bound states with zero energy -- one-dimensional flat band, which was first discussed by Kopnin and Salomaa in 1991 \cite{KopninSalomaa1991}. The end points of this flat band are determined by projections of the Weyl points to the direction of the vortex axis. Due to bulk-surface correspondence, the surface of $^3$He-A and of 3D topological semimetals contains another exotic object -- the Fermi arc --  the 1D Fermi line in the 2D momentum space, which terminates on two Weyl points (Fig. \ref{classes} {\it bottom right}).  The flat band in the vortex core and the Fermi arc on the surface are analogs of the Dirac string terminating on two magnetic monopoles.

\begin{figure}
  \includegraphics[width=1.0\textwidth]{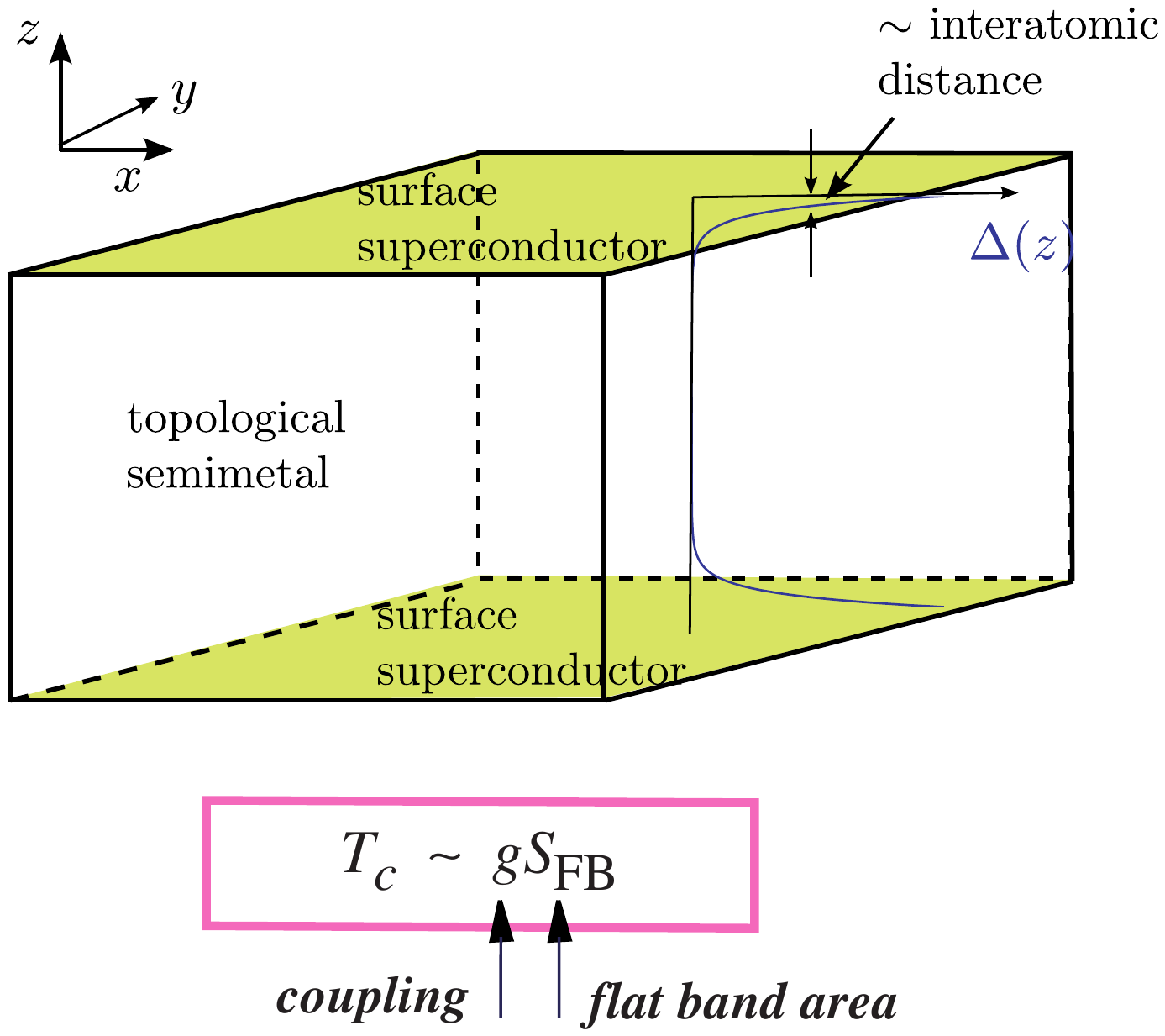}
\caption{Superconductivity formed on the surface of a system which has nodal line in bulk and thus the flat band on the surface. Due to the singular DoS $\nu(\epsilon)\sim \epsilon^{-1}$ of the surface flat band, superconductivity emerges on the surface earlier than in the bulk material. The critical temperature  $T_c$ of the surface superconductivity depends linearly on the coupling $g$, which should be contrasted with the exponential suppression of transition temperature in the bulk superconductors}
\label{SC}       
\end{figure}

Another important class of gapless topological systems contains 3D semimetals and superconductors with topologically protected lines of zeroes in momentum space. For systems with nodal lines the bulk-surface correspondence gives rise to the 2D flat band on the surface of material of this class or at the twin boundaries -- all electrons within this band have zero energy. The flat band spectrum terminates by the line obtained by projection of the nodal line to the plane of the surface, see  Fig. \ref{SpiralFlatBand}. Flat band has an extremely singular density of states, and this property of systems with flat band is very important. In particular, it gives the linear dependence of the critical temperature of superconducting transition on the coupling: $T_c\sim gS_{\rm FB}$, where $S_{\rm FB}$ is the area of the flat band in momentum space. This should be contrasted with the exponential suppression of transition temperature in bulk superconductors.  Flat band may give rise to surface or interface superconductivity with high transition temperature \cite{Heikkila2011,KopninHeikkilaVolovik2011,Kopnin2011} (see Fig. \ref{SC}) and may open the route to room-temperature superconductivity. 

In conclusion, the momentum space topology became the main tool for investigation of the robust properties
of fermionic condensed matter systems and exotic gapless fermions, including Weyl fermions, Majorana fermions (see recent review \cite{Alicea2012}), Fermi arc and flat band. It is also applicable for investigations of the topologically nontrivial vacua in relativistic quantum field theories, including quantum chromodynamics, see Refs. \cite{Volovik2011,ZubkovVolovik2012}  and references therein.  The highly degenerate topologically protected state -- the flat band --  is a generic phenomenon.  The classes of topological matter, which experience the flat band, are waiting for its exploration. 
 To reach the room-temperature superconductivity we must search for or artificially  create the systems
 which experience the non-topological flat band in bulk or topologically protected flat bands on the surface or at the interfaces. In the latter case the proper arrangement of many twins or grain boundaries
 is needed to obtain the bulk superconductivity with high temperature. Existence of localized superconducting domains at elevated
temperatures has been suggested  in Refs. \cite{SilvaTorresKopelevich2001,LukyanchukKopelevich2009}.

\begin{acknowledgements}
 This work
is supported in part by the Academy of Finland and its COE program

\end{acknowledgements}



\end{document}